\documentclass[twocolumn]{article}
\usepackage[accepted]{aistats_arxiv}

\usepackage{amsfonts}
\usepackage{color}
\usepackage[usenames,dvipsnames]{xcolor}
\usepackage{multirow}
\usepackage{enumitem}
\usepackage[linesnumbered, titlenumbered,ruled]{algorithm2e}

\SetCommentSty{mycommfont}

\usepackage[many]{tcolorbox}
\usepackage{empheq}
\usepackage{xr-hyper} 
\usepackage{hyperref}
\usepackage[nameinlink, capitalise]{cleveref}
\hypersetup{colorlinks,linkcolor={blue},citecolor={blue},urlcolor={magenta}}  

\usepackage{float}
\usepackage{bbm}

\usepackage{multicol}
\usepackage{amsmath}
\usepackage{graphicx}
\usepackage{caption}
\usepackage{subcaption}

\newtheorem{thm}{Theorem}
\newtheorem{lem}[thm]{Lemma}

\newcommand{\tpi}{\tilde \pi}
\newcommand{\rmoff}{\mathrm{off}}

\newcommand{\rmtry}{\mathrm{try}}
\newcommand{\rmut}{\mathrm{ut}}
\newcommand{\rmmax}{\mathrm{max}}
\newcommand{\rmmin}{\mathrm{min}}

\makeatletter
\newcommand{\removelatexerror}{\let\@latex@error\@gobble}
\makeatother

\usepackage{float}
\usepackage{xr}

\usepackage[round]{natbib}

\usepackage{xr}
\usepackage{appendix}
\makeatletter
\newcommand*{\addFileDependency}[1]{
  \typeout{(#1)}
  \@addtofilelist{#1}
  \IfFileExists{#1}{}{\typeout{No file #1.}}
}
\makeatother
\begin{document}
\twocolumn[
\aistatstitle{ATLAS: Adapting Trajectory Lengths and Step-Size for Hamiltonian Monte Carlo}
\aistatsauthor{ Chirag Modi}
\aistatsaddress{Center for Computational Mathematics, Flatiron Institute\\
Center for Computational Astrophysics, Flatiron Institute\\
Center for Cosmology and Particle Physics, New York University  
} 
]

\begin{abstract}

Hamiltonian Monte-Carlo (HMC) and its auto-tuned variant, the No U-Turn Sampler (NUTS) can struggle to accurately sample distributions with complex geometries, e.g., varying curvature, due to their constant step size for leapfrog integration and fixed mass matrix.
In this work, we develop a strategy to locally adapt the step size parameter of HMC at every iteration by evaluating a low-rank approximation of the local Hessian and estimating its largest eigenvalue. 
We combine it with a strategy to similarly adapt the trajectory length by monitoring the no U-turn condition, resulting in an adaptive sampler, ATLAS:~\emph{adapting trajectory length and step-size}\footnote{Python implementation of ATLAS is available 
at \href{https://github.com/modichirag/AtlasSampler}{https://github.com/modichirag/AtlasSampler}.}.
We further use a delayed rejection framework for making multiple proposals that improves the computational efficiency of ATLAS, and develop an approach for automatically tuning its hyperparameters during warmup.  
We compare ATLAS with 
NUTS on a suite of synthetic and real world examples, and show that i) unlike NUTS, ATLAS is able to accurately sample difficult distributions with complex geometries, ii) it is computationally competitive to NUTS for simpler distributions, and iii) it is more robust to the tuning of hyperparamters. 
\end{abstract}

\section{Introduction}
\label{sec:introduction}

Hamiltonian Monte Carlo (HMC) is one of the most popular Markov Chain Monte Carlo (MCMC) algorithms for Bayesian inference in high dimensions~\citep{duane1987hybrid, neal2011mcmc, betancourt2017conceptual}.
However, its performance is sensitive to the tuning of the algorithm parameters--- the step-size of leapfrog integration, trajectory length (integration time), and the mass matrix (pre-conditioner). 
Thus development of auto-tuned variants like the No-U-Turn Sampler (NUTS)~\citep{hoffman2014no} that locally adapt trajectory length has been crucial for its widespread usage in probabilistic programming~languages. 

\begin{figure}
    \centering
    \includegraphics[width=1.0\columnwidth]{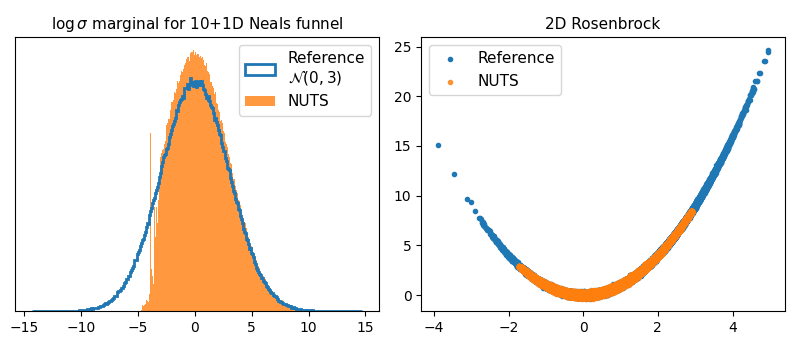}    
    \caption{NUTS with constant step-size can fail to accurately sample distributions with varying curvature like Neal's funnel (left) and Rosenbrock (right).}
    \label{fig:nutsfail}
\vspace{-17pt}
\end{figure}

There are however no widely used samplers that similarly adapt the step-size parameter of HMC. Most variants, including NUTS, tune a step-size in an initial warm-up phase, generally by targeting an acceptance rate~\citep{hoffman2014no}, and then fix it during sampling.
MCMC algorithms with constant step size and mass matrix can struggle to accurately sample distributions with complex geometries like strongly varying curvature, e.g.,  Neal's funnel and Rosenbrock, which arise in hierarchical models~\citep{betancourt2015hamiltonian}.
Regions with high curvature require small step sizes for numerical stability, while flat regions with large volume need large step sizes for efficient sampling. 
Hence, as shown in \cref{fig:nutsfail}, algorithms like NUTS with fixed step size fail to sample these distributions. 

Recent works like \cite{autoMALA} and \cite{Bou-Rabee24b} have explored strategies for adapting step size by doubling or halving the step size to bound the energy error (difference of log-probabilities) in leapfrog integration.
While this allows one to navigate high curvature regions, it becomes computationally inefficient for simpler distributions. 

\cite{Modi_2023} laid out a different approach for step size adaptation by introducing a \textit{delayed rejection} (DR) approach to HMC.
In this, upon a rejection in the MCMC chain, additional proposals are made by successively reducing the step size until a proposal is accepted, or a maximum number of proposals are made~\citep{tierney1999some, green2001delayed}.
\cite{Turok24} extended this to generalized HMC with partial momentum refresh~\cite{horowitz1991generalized}.
However DR framework introduces new hyperparameters to be tuned, e.g., the reduction factor of the step size for successive proposals.
The step-size adaptation is also limited to a pre-selected sequence of step sizes, and is not informed by the local geometry.

In this work, we develop a new strategy for automatically and continuously adapting the step size of HMC using the local geometry of the distribution.
The main idea is to use the positions and gradients along a HMC trajectory to approximate the local Hessian and its maximum eigenvalue which, for quadratic potentials, bounds the maximum stable step size for leapfrog integration~\citep{Leimkuhler04, Hoffman22}.
Hence it can be used to construct a distribution of locally feasible step sizes from which we sample the step size for the next HMC trajectory. 

We combine this approach with a strategy for adapting trajectory lengths introduced in \cite{Bou-Rabee24} to build a new adaptive sampler, ATLAS- \textit{adapting trajectory length and step-size}. 
To improve the computational efficiency of ATLAS, we also implement it in a delayed rejection framework that adapts step size only when necessary~\citep{Modi_2023}.
We will show that ATLAS meets the following desiderata:
\begin{itemize}[itemsep=0.5pt, topsep=0.5pt]
    \item accurate: unlike NUTS, it correctly samples distributions with complex geometries.
    \item computationally efficient: for simpler distributions without these pathalogies, ATLAS is competitive to state-of-the-art samplers like NUTS.
    \item automatic parameter tuning: all fixed parameters are tuned automatically in the warmup stage.
    \item robust: unlike HMC, the accuracy is broadly insensitive to the choices of hyperparameters and the parameters tuned during the warmup.
\end{itemize}

With these goals, the paper is organized as follows. We begin by reviewing the basics of MCMC and HMC, as well as summarizing the delayed rejection framework and the strategy for trajectory adaptation in \cref{sec:background}.
We describe the strategy for adapting the step size in \cref{sec:stepadapt} and present the ATLAS sampler in \cref{sec:atlas}.
In \cref{sec:experiments}, we show numerical experiments to compare ATLAS with NUTS. We conclude in~\cref{sec:discussion}.
\section{Background}
\label{sec:background}

We seek to generate samples from a differentiable and (generally) unnormalized target probability density function $\pi(\theta)$ with parameters $\theta \in \mathbb{R}^D$. 

\subsection{Metropolis Hastings (MH) algorithm}
MCMC methods generate samples from the target distribution by constructing a Markov Chain defined by transition kernel $k(\theta, \theta')$, which gives the probability of transitioning from current state $\theta$ to the next state $\theta'$.
MH algorithms propose a new state $\theta'$ from a proposal distribution $q(\theta, \theta')$ and accept it with probability $\alpha(\theta, \theta')$. If rejected, the next state remains as $\theta$. Thus the transition kernel (upto a rejection term that can be ignored for simplicity as it is always symmetric) is
$ k(\theta, \theta') = q(\theta, \theta') \alpha(\theta,\theta') $.

MCMC chain will sample the correct target distribution if it is invariant under the transition kernel. This is achieved by maintaining detailed balance~(DB). 
\vspace{-2pt}
\begin{equation}
    \pi(\theta) k(\theta, \theta') = \pi(\theta') k (\theta', \theta)
    \label{eq:db}
\vspace{-2pt}
\end{equation}
This allows us to evaluate the acceptance probability 
 $\alpha(\theta, \theta') = 1 \wedge \frac{\pi(\theta')\, q(\theta', \theta)}{\pi(\theta)\, q(\theta, \theta')}$, where  $a\wedge b = \mathrm{min(a, b)}$.

\subsection{Hamiltonian Monte Carlo (HMC)}

HMC uses gradient information to \textit{efficiently} explore the target density $\pi$. HMC treats $\theta$ as position vector and introduces an auxiliary momentum vector $\rho$ for a fictitious particle with mass matrix $M$. This corresponds to a physical system with  Hamiltonian $H(\theta, \rho) = \frac{1}{2} \rho^T M^{-1} \rho - \log \pi (\theta)$ and the target density: $\tilde{\pi}(\theta, \rho) \propto \pi(\theta) \times \text{normal}(\rho \mid 0, M)$. Henceforth we will use $\tilde \pi$ to denote this Gibbs density which is the combined density of a state in phase space $x = (\theta, \rho)$. HMC samples this target Gibbs density, the $x-$marginal of which is the target distribution $\pi$.

\paragraph{Proposal Map:} HMC simulate a Markov chain in two steps: 1) resample the momentum $\rho \sim \textrm{normal}(0, M)$ as a Gibbs update to get the current state $x=(\theta, \rho)$. 2) Perform a MH update by simulating the Hamiltonian dynamics using a leapfrog integrator ($\mathcal{L}^n_\epsilon$) for $n$ steps with step size $\epsilon$, followed by a momentum flip (negation, $\mathcal{F}$) to reach the proposal point $y = (\theta', -\rho') = \mathcal{F}\mathcal{L}^n_\epsilon(x)$.
For a given step-size and number of leapfrog steps, HMC proposal map is \emph{deterministic}, volume preserving, and an involution i.e. it is time reversible: $(\mathcal{F}\mathcal{L}^n_\epsilon)^{-1} = \mathcal{F}\mathcal{L}^n_\epsilon$.
Hence the transition kernel for HMC up-to a rejection term is $k(x, y) = \tpi(x) \delta_D(y - \mathcal{F}\mathcal{L}^n_\epsilon(x))$ where $\delta_D$ is the dirac-delta function. Imposing DB with \cref{eq:db} gives the acceptance probability $\alpha(x, y) = \frac{\tpi(y)}{\tpi(x)}$.
\begin{figure}
\centering
\includegraphics[width=0.75\columnwidth]{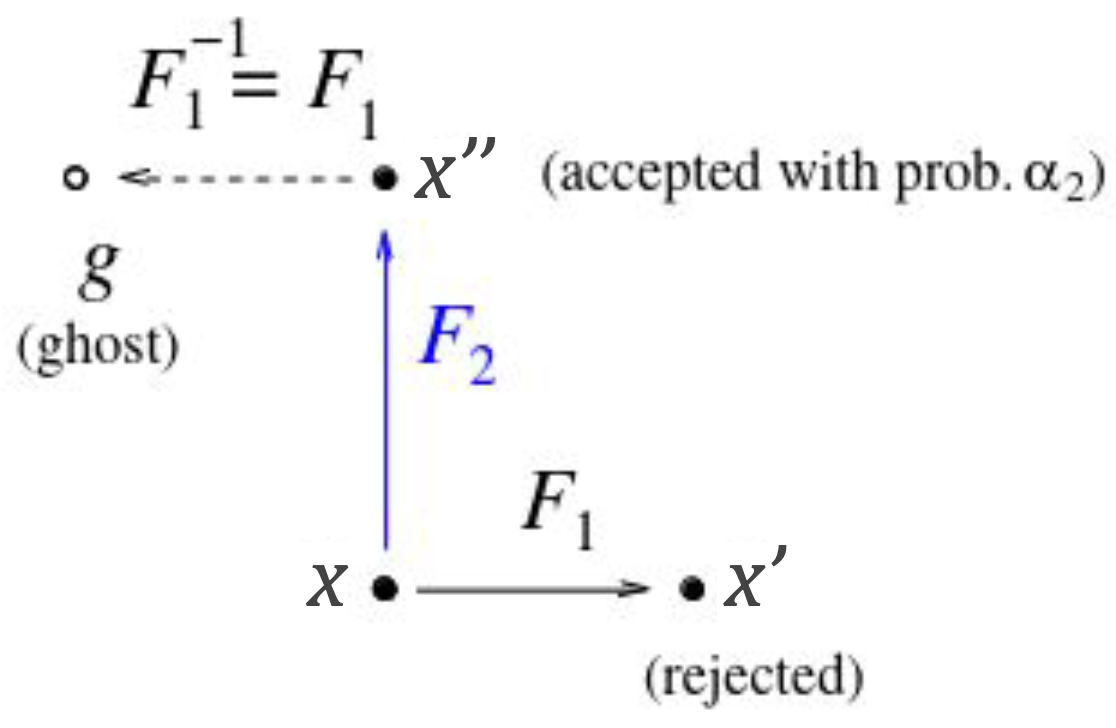}
\caption{DR-HMC schematic adapted from \cite{Modi_2023} with current state $x$, proposed state $x''$, previously rejected proposals $x'$, and ghost proposal $g$.}
\label{fig:drhmc}
\vspace{-7pt}
\end{figure}

\vspace{-7pt}
\subsection{Delayed rejection (DR) approaches}
Since ATLAS will be implemented in a DR framework for computational efficiency, we present its brief summary. We discuss it in context of random walk MH algorithm~\citep{tierney1999some, green2001delayed} as it is more instructive. 

The main idea is that if a proposal in the MCMC chain is rejected, 
instead of falling back to the original state, DR methods make additional proposals in the \textit{same} iteration using a \textit{different} proposal kernel.
Let the first and second proposal maps be $F_1, F_2$ and corresponding proposal distribution $q_1(x, x'), q_2(x, x')$.
The proposal kernel upto a rejection term is then 
\begin{align}
\vspace{-5pt}
k(x,y) &= q_1(x,y)\alpha_1(x,y) \nonumber +\\
&\int q_1(x,s) [1-\alpha_1(x,s)]q_2(x,y)\alpha_2(x,y)ds    
\label{eq:dr}
\vspace{-5pt}
\end{align}
The acceptance probability of the first proposal is the same as standard MH: $ \alpha_1(x, x') = 1 \wedge \frac{\tilde\pi(x')\, q(x', x)}{\tilde\pi(x)\, q(x, x')}$.
However, to maintain DB, the acceptance probability for second proposal accounts for the fact that we have made, and rejected, a first proposal. It is given by
\begin{equation}
        \alpha_2(x, x'') = 1 \wedge \left( 
            \frac{\tilde\pi(x'')\, q_2(x'', x)}{\tilde\pi(x)\, q_2(x, x'')}             \frac{q_1(x'', x^g) [1-\alpha_1(x'', x^g)]}
          { q_1(x, x') [1-\alpha_1(x,  x')]} 
          \right)
    \label{eq:almhdr}
\end{equation}
where $x^g$ is the \textit{ghost} proposal that needs to be additionally evaluated- this is the \emph{first proposal that a (hypothetical) reversed MCMC chain starting from $x''$ would have made and rejected}. A schematic for a full MCMC iteration with DR is shown in \cref{fig:drhmc}. 

\subsection{Trajectory Adaptation with NoUT}
\label{sec:nout}

\begin{figure}
    \centering
    \includegraphics[width=\columnwidth]{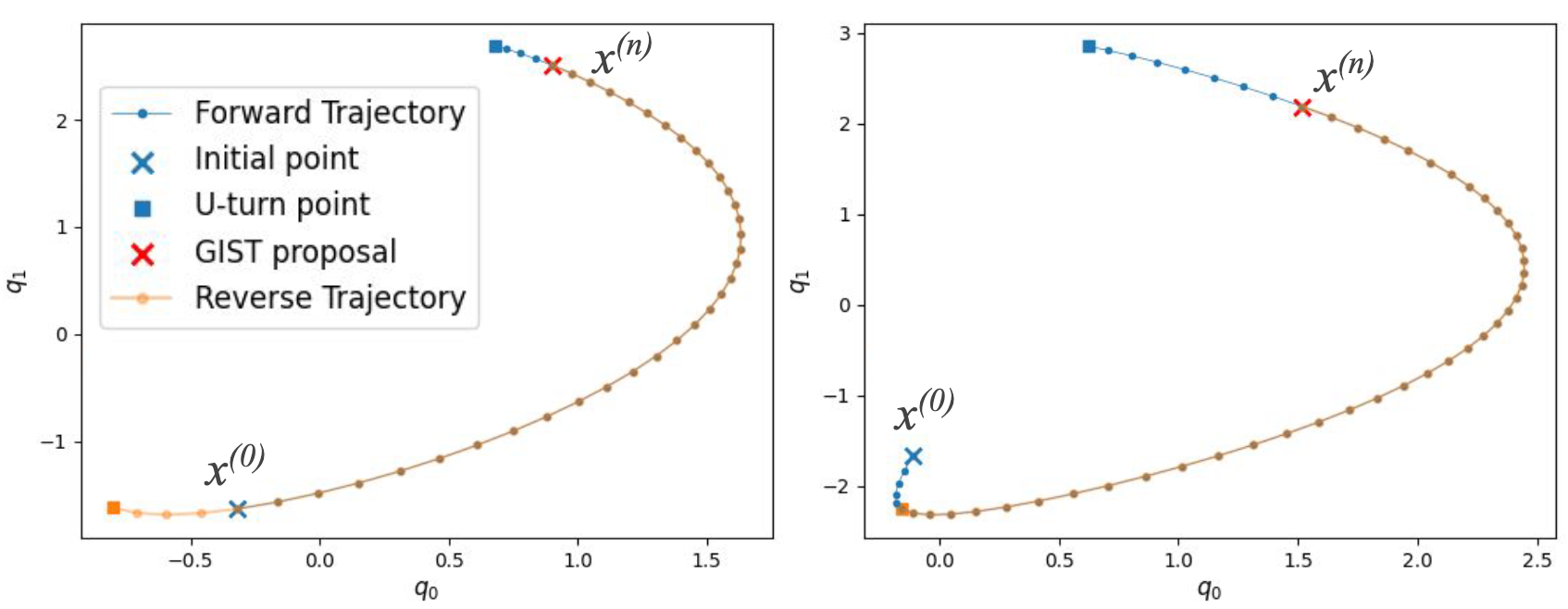}           
    \caption{(Left) One iteration for NoUT sampler from the initial point $x^{(0)}$, and proposal $x^{(n)}$. (Right) A \textit{sub u-turn}: reverse trajectory (orange) ends before reaching the initial point $x^{(0)}$, leading to rejection.}
    \label{fig:gist}
\vspace{-10pt}
\end{figure}

While our step-size adaptation approach can be combined with NUTS, here we use another strategy introduced in \cite{Bou-Rabee24} due to its simplicity.
The main idea however is same as NUTS- we continue on a trajectory until a u-turn is made. Hence we call this algorithm NoUT (No U-Turn). 
A u-turn occurs when the distance of the next position in the trajectory from the initial position is less than that of the current position, i.e., $||\theta^{(j+1)} - \theta^{(0)} ||_2 < ||\theta^{(j)} - \theta^{(0)} ||_2$
where $||\cdot||_2$ is the L2 norm and $x^{(0)} = (\theta^{(0)}, \rho^{(0)})$ is the initial state. 
When this condition is met, we stop the trajectory and set the u-turn trajectory length:~$n_\rmut = j$. 

A proposal $y=x^{(n)}$ is made from states on the trajectory $[x^{(0)},...,x^{(n_\rmut)}]$ according to a distribution which depends only on $n_\rmut$ i.e., $q(x, y) = q(n|x^{(0)}) = q(n|n_\rmut)$.
For detailed balance, we also need to evaluate the probability of choosing $x^{(0)}$ from $x^{(n)}$, i.e. $q(n|x^{(n)})$, for which a \textit{ghost} trajectory is simulated in the reverse direction starting from $x^{(n)}$. A schematic of this is in the left panel of \cref{fig:gist}.
The transition kernel for NoUT approach is $k(x,y)=q(n|x)\delta_D(y-\mathcal{FL}_\epsilon^n(x))$ and the acceptance probability to maintain DB is:
\begin{equation}
\alpha(x, y) = \alpha(x^{(0)}, x^{(n)}) = 1 \wedge \frac{\tilde\pi(x^{(n)})\, q(n | x^{(n)})}{\tilde\pi(x^{(0)})\, q(n | x^{(0)})}
\label{eq:algist}    
\end{equation}
For the proposal distribution $q(n|n_\rmut)$, we use discrete-uniform distribution parameterized with $f_\rmoff \in [0, 1]$, i.e., we sample a $n \sim \mathcal{U}(\lfloor{n_\rmut f_\rmoff}\rfloor, n_\rmut)$, equivalently $\theta^{(n)} \sim \mathcal{U}(x^{(\lfloor{n_\rmut f_\rmoff}\rfloor)},...,x^{(n_\rmut)})$.
Large value of $f_\rmoff$ is desirable as it results in larger jumps per acceptance.
However it also increases the chance of rejection due to the possibility that the reverse trajectory from $x^{(n)}$ makes a U-turn before $x^{(0)}$ is reached in the reverse trajectory (see right panel of \cref{fig:gist} for illustration). Then $q(n|x^{(n)}) = 0$ and $x^{(n)}$ gets rejected. We refer to this as a \textit{sub-u-turn}.
We will randomly vary $f_\rmoff$ between 0.33 and 0.66 at every iteration to strike a balance between these trade-offs.
We give full algorithm for NoUT strategy in \cref{sec:nout_details}.

\section{Step-size adaptation}
\label{sec:stepadapt}

We now present our approach to adapt step size for HMC. Broadly, we will construct a local distribution of the largest feasible step-sizes  $q(\epsilon | x)$ at every iteration and sample the step size for the next trajectory from it.  
For quadratic potentials, $V = \frac{1}{2}\theta^T H \theta$, the largest stable step-size for the leapfrog integrator is $\epsilon \leq \frac{2}{\sqrt{\lambda_\rmmax}}$ where $\lambda_\rmmax$ is the largest eigenvalue of the Hessian $H$ \citep{Leimkuhler04}. 
Hence we will approximate this eigenvalue and use it to construct $q(\epsilon|x)$. The full algorithm is given in \cref{sec:stepadapt_details}.

\vspace{-5pt}
\paragraph{Hessian approximation:}
We use the LBFGS algorithm to approximate the local Hessian. 
This combines the position and gradient information to obtain a low-rank approximation of Hessian using only vector products, achieving $\mathcal{O}(D^2)$ complexity that can scale to high dimensions. 
One way to obtain the positions and gradients for LBFGS at the beginning of each iteration is by running a small HMC trajectory of length $N_H(=10)$ with a baseline step size $\epsilon_0$. 
However as we do later in DR framework, we can instead use positions and gradients from the rejected trajectory of the first proposal and avoid any extra computations.

We also experimented with approximation to Hessian used in \cite{Cai24bam} for variational inference, and Gauss Newton approximation~\citep{Hoffman22}. While the former performed similarly to LBFGS, the latter failed as it evaluates an \emph{expectation}, resulting in poor approximation of local Hessian.  

\vspace{-5pt}
\paragraph{Stable step-size: } We approximate the maximum eigenvalue $\lambda_\rmmax$ of the Hessian using power iteration. This gives an estimate for the stable step size $\epsilon_s = \frac{1}{2\sqrt{\lambda_\rmmax}}$. We have used the factor $\frac{1}{2}$ instead of $2$ to allow some margin in constructing the step size~function.





\vspace{-5pt}
\paragraph{Step-size distribution:}
To construct the distribution $q(\epsilon|x)$, we use a lognormal distribution $\log\mathcal{N}(\mu^*, \sigma^*)$. The mean is set to the estimated stable step-size $\mu^*=\epsilon_s$. We experimented with different values for $\sigma*$, finding $\sigma^*=1.2$ to empirically perform the best.
The corresponding normal distribution to this lognormal is: $\log \epsilon \sim \mathcal{N}\big(\log \mu^* -\frac{(\log \sigma^*)^2}{2}, \log \sigma^*\big)$.
We also experimented with distributions like scaled-beta distribution which are bounded, but found them to be more expensive than lognormal.

\vspace{-5pt}
\paragraph{Acceptance probability:}
We draw a step size $\epsilon\sim q(\epsilon|x)$ with which to simulate the next trajectory. 
Given that the Hamiltonian trajectory ($\mathcal{FL}^n_{\epsilon_x}$) is deterministic for a step size and the number of leapfrog steps, the transition kernel is $k(x, y) = \tpi(x) q(\epsilon|x) \delta(y - \mathcal{FL}_\epsilon^n(x))$.
Thus the acceptance probability is:
\begin{equation}
    \alpha(x, y) = 1 \wedge \frac{\tilde\pi(y)\, q(\epsilon | y)}{\tilde\pi(x)\, q(\epsilon| x)}
    \label{eq:mhstep}
\end{equation}
Same as DR and NoUT proposals, $q(\epsilon | y)$ needs to be constructed at the proposal ($y$) to evaluate \cref{eq:mhstep}.

\section{ATLAS}
\label{sec:atlas}

We now construct ATLAS sampler by combining step-size adaptation with NoUT trajectory~adaptation. 

\subsection{Simple adaptive sampler}
\label{sec:simpleatlas}

A straightforward strategy to combine the two adaptations would be the following: at every iteration,
(1) construct the step-size distribution and propose a step-size $\epsilon$; then
(2) make a NoUT proposal $y=x^{(n)}$ by simulating a trajectory with this step-size until u-turn. The transition kernel for this algorithm would be 
$k(x, y) = \tpi(x) q(\epsilon|x) q(n|x) \delta(y - \mathcal{FL}_\epsilon^n(x))$ and the corresponding acceptance probability is \begin{equation}
    \alpha(x, y) = 1 \wedge \frac{\tilde\pi(y)\, q(\epsilon | y)\, q(n | y)}{\tilde\pi(x)\, q(\epsilon | x)\, \, q(n| x)}
    \label{eq:mhatlassimple}
\end{equation}
To evaluate this, we will need to construct the step-size distribution $q(\epsilon|y)$ at $y=x^{(n)}$, and simulate a reverse trajectory with~$\epsilon$ to construct $q(n|y)$.

While this algorithm is adaptive and accurately samples difficult  distributions, we show in \cref{sec:atlas_simple_details} that it is expensive for simple distributions. This is for two reasons.  
1) Approximating local Hessian requires extra computations of running short HMC trajectory. This might not be useful at every iteration. 
2)~NoUT trajectories can lead to rejections due to sub-u-turns (\cref{fig:gist}, right panel). 
If we happen to sample small step size for such trajectories, the large number of leapfrog steps before u-turn is a lot of wasted~computation. 

\subsection{Adaptivity in DR framework}
\label{sec:dratlas}

It is not necessary to adapt step-size at every iteration, and a well chosen baseline step size $\epsilon_0$ can be adequate in a large region of phase space. 
Then we can make our adaptive sampler computationally efficient by using DR framework.
In this, we make the first proposal as a NoUT proposal ($x'=x^{(n_1)}$) from a trajectory simulated with step size $\epsilon_0$.
If this gets accepted, we continue to the next iteration. 
Otherwise we adapt the step size \textit{using the rejected trajectory to approximate Hessian} and make a delayed~proposal.

\paragraph{Delayed proposal for trajectory adaptation:}
Hastings correction for delayed proposal requires us to evaluate the probability of rejecting the ghost proposal ($1-\alpha_1(x'', x^g)$) made from second proposal $x''$. For NoUT as the first proposal map, there are many possible ghost states depending on the value $n_g \sim q(n|x'')$.
Hence evaluating $1-\alpha_1(x'', x^g)$ will require marginalizing over all these states which is computationally expensive.
This is avoided if we set $n_2$, the number of leapfrog steps to the second proposal, an invertible function of $n_1$. 
Then  we only need to evaluate the rejection probability for $n_g = n_1$ as any other value will make the reverse transition $x'' \rightarrow x$ impossible.

While this solves the issue of marginalizing over many ghost states, it assumes that the trajectory simulated for the first proposal is reliable.
In the regions of high curvature where the baseline step size $\epsilon_0$ is unstable, $n_1$ might be 0 and this assumption breaks down.
We refer to this scenario as \textit{DR-upon-failure} i.e. when the NoUT fails to generate a trajectory to inform delayed proposals.
It can be identified easily, e.g., when $n_\rmut < n_\rmmin$, i.e., the u-turn length is less than a minimum number of steps, say $n_\rmmin=3$.
We treat delayed proposal for this scenario separately and sample the trajectory length from a global distribution of trajectory lengths $q_g(n)$ constructed during the warmup.

\subsubsection{ATLAS Algorithm}
\label{sec:atlasalgo}
Now we develop an adaptive algorithm in a delayed rejection framework. The pseudocode of a single iteration is in \cref{alg:atlas}. A schematic flowchart is in \cref{fig:flowchart}. We describe it in words here-

\noindent Starting at initial state $x$, run a trajectory upto u-turn (say $n_\rmut$ steps) with step size $\epsilon_0$. This is map $F_1$ of DR.

\begin{enumerate}[topsep=1pt]
    \item\textbf{If} $n_\rmut > n_\rmmin$: 
         Make a NoUT proposal: $n_1\sim q(n|x)$, $x'=x^{(n_1)}$. 
        Accept/reject with \cref{eq:algist}.
        If accepted, set $y=x'$ and end iteration. \textbf{Else:}
        \begin{enumerate}
            \item \label{it:subuturn} \textbf{If} rejected due to sub-u-turn, no delayed proposal\footnote{This choice is made  for computational efficiency. Adapting step size is likely not going to address reasons that lead to sub-u-turn rejections. Empirically, we find that not making DR proposals here reduces computational cost without affecting the accuracy and robustness.}. Set $y=x$ and end iteration.
            
            \item  \textbf{Else} make delayed proposal with map $F_2$. Use states from rejected NoUT trajectory to adapt step size $\epsilon_2$. Set trajectory length equal to first proposal, $n_2=\lfloor \epsilon_0 n_1/\epsilon_2\rfloor$). Propose $x'' = F_2(x) = \mathcal{FL}^{n_2}_{\epsilon_2}(x)$. To evaluate acceptance probability of \cref{eq:almhdr}, simulate a ghost NoUT trajectory from $x''$ with $\epsilon_0$. Evaluate rejection probability for $x^g$ at $n_1$ with \cref{eq:algist}. Ensure other conditions are met for delayed proposal, i.e., i) $n^g_\rmut > n_\rmmin$, ii) no sub u-turn in ghost trajectory, and iii) $q(n_1|x'') \neq 0$. Evaluate MH criterion and accept/reject the proposal $x''$. If rejected, fall back to $x$.
        \end{enumerate}
    \item \textbf{Else} ($n_\rmut \leq n_\rmmin$): make DR-upon-failure proposal. Adapt the step size $\epsilon_3$ by simulating small HMC trajectory\footnote{Starting at step size=$\epsilon_0$, we reduce it by a factor of 2 until we simulate a trajectory of length $N_H(\sim 10$, say).}. Sample trajectory length $n_3\sim q_g(n)$ and propose $x''' = \mathcal{L}^{n_3}_{\epsilon_3}(x)$. Simulate a ghost trajectory from $x'''$ with $\epsilon_0$ to ensure $n^g_\rmut \leq n_\rmmin$. Evaluate MH criterion with probability \cref{eq:mhstep} and accept/reject $x'''$. If rejected, fall back to $x$.
\end{enumerate}

\vspace{-5pt}
\begingroup
\removelatexerror
\begin{algorithm}[H]
\DontPrintSemicolon
\SetAlgoNoEnd
    \SetKwInOut{KwIn}{Input}
    \SetKwInOut{KwOut}{Output}
    \SetKwComment{Comment}{$\triangleright$\ }{}
    \SetKwFunction{TrajectoryUptoUTurn}{TrajUptoUTurn}
    \SetKwFunction{GISTProposal}{GISTProposal}
    \SetKwFunction{StepSizeAdaptation}{StepSizeDist}
    \SetKwFunction{DelayedStepUponFailure}{DelayedStepUponFailure}
    \SetKwFunction{HMCTraj}{HMCTraj}
    \SetKwFunction{LeapFrog}{LeapFrog}
    \SetKwFunction{StepSizeAdaptationwithHMC}{StepSizeDistHMC}
    \SetKwProg{Fn}{Function}{:}{}

    \KwIn{Target $\tpi$, Current state $x$, 
    Offset $f_\rmoff$,
    Baseline step size $\epsilon_0$, 
    Distribution of trajectories $q_g(n)$, 
    Min. number of leapfrog steps $n_\rmmin=3$,
    Samples for approximating Hessian $N_H=10$,
    }
    \KwOut{Next state in the ATLAS trajectory}
    
    \vspace{\baselineskip}
    $\tilde{x}, \tilde{g}, n_\rmut$ = \TrajectoryUptoUTurn{$x, \epsilon_0$} \tcp{$\tilde{}$ is list}
    \uIf(\hfill \tcp*[h]{Reliable NoUT traj.}){$n_{\rmut} > n_\rmmin$}
    {   
        $n_1\sim q(n|x) = \mathcal{U}(\lfloor f_\rmoff n_\rmut \rfloor, n_\rmut)$ \;  
        $x' = \tilde{x}^{(n_1)}$ \tcp{NoUT as first proposal} 
        $\alpha_1 = \frac{\tilde \pi(x')q(n|x')}{\tilde \pi(x)q(n|x)}$; 
        $\, u \sim \mathcal{U}(0, 1)$\;
        \lIf{$u \leq \alpha_1$}
        {
            \Return $x'$
        }
        \lElseIf(\hfill \tcp*[h]{No DR if 1$^{st}$ proposal rejected due to sub-U-turn}){$q(n_1|x') = 0$ } {\Return $x$}
        \Else{ 
                $q(\epsilon|x)$ = \StepSizeAdaptation{$\tilde{x}, \tilde{g}$} \; 
                $\epsilon_2 \sim q(\epsilon|x)$;
                $\, n_2 = \lfloor\epsilon_0 n_1 / \epsilon_2 \rfloor$ \;
                
                $x'' = \mathcal{FL}_{\epsilon_2}^{n_2}(x)$ \tcp*{Second proposal}                
                \tcc{Ghost trajectory, has to be $>n_1$}
                $\tilde{x_g}, \tilde{g_g}, n''_\rmut$ = \TrajectoryUptoUTurn{$x'', \epsilon_0$};
                $x^g = \tilde{x_g}^{(n_1)}$\;
                \lIf{$n''_{\rmut} < n_\rmmin$ or $q(n_1|x'')=0$ or $q(n_1|x^g)=0$ }
                { \Return $x$ }
                    $ q(\epsilon|x'')$ ~=~\StepSizeAdaptation{$\tilde{x_g}, \tilde{g_g}$}\;
                    $\alpha_g = \frac{\tilde\pi(x^g)q(n_1|x^g)}{\tilde\pi(x'')q(n_1|x'')}$\;
                    
                    $\alpha_2 = \frac{\tilde \pi(x'') q(\epsilon_2|x'')q(n_1|x'') (1-\alpha_g)}{\tilde\pi(x) q(\epsilon_2|x)q(n_1|x)(1-\alpha_1)}$;    
                    $u \sim \mathcal{U}(0, 1)$\;
                    \lIf{$u \leq \alpha_2$}
                    {\Return $x''$}
                    \lElse{\Return $x$}    
            }
        }    
    \Else(\tcp*[h]{DR-upon-failure.}){
        $q(\epsilon|x)$ = \StepSizeAdaptationwithHMC{$x$} \; 
        $\epsilon_3 \sim q(\epsilon|x)$; $n_3\sim q_g(n)$; $x'''=\mathcal{L}_{\epsilon_3}^{n_3}(x)$ \;
        \tcc{Check DR-upon-failure for ghost}
        $\tilde{x'''}, \tilde{g'''}, n'''_\rmut$ = \TrajectoryUptoUTurn{$x''', \epsilon_0$} \;
        \lIf{$n'''_\rmut > n_\rmmin$}{\Return $x$}
        \uElse{
        $q(\epsilon|x''')$ = \StepSizeAdaptationwithHMC{$x'$} \; 
        $\alpha_3 = \frac{\tpi(x''') q(\epsilon_3|x''')}{\tpi(x) q(\epsilon_3|x)}$; $u \sim \mathcal{U}(0, 1)$\;
        \lIf{$u \leq \alpha_3$}{\Return $x'''$}
        \lElse{\Return $x$}
        }
    }
    \vspace{\baselineskip}
    \Fn{\StepSizeAdaptationwithHMC{$x, \epsilon = \epsilon_0$}}{
        \Repeat(\tcp*[h]{Goal: simulate HMC of $N_H$ steps}){unable to simulate traj. of length $N_H$}{
        $\epsilon = \epsilon/2$ \tcp*[h]{Try by reducing stepsize} \; 
        $\tilde{x}, \tilde{g}$ = \HMCTraj{x, $\epsilon, N_H$} \;
        }
        \textbf{return} \StepSizeAdaptation{$\tilde{x}, \tilde{g}$}
    }
    \caption{ATLAS}
    \label{alg:atlas}
\end{algorithm}
\vspace{-2pt}
\endgroup

The kernel for ATLAS (upto rejection) is:
\vspace{-3pt}
\begin{align}
    k&(x, y)= \mathbbm{1}(n_\rmut > n_\rmmin)\Big[q(n|x) \alpha_1(x, y) +\nonumber  \\
    &\big(1-\mathbbm{1}(subuturn)\big) q(\epsilon|x)q(n|x) [1-\alpha_1(x, y)] \alpha_2(x, y)\Big] + \nonumber \\
    &\big(1 - \mathbbm{1}(n_\rmut > n_\rmmin)\big)q(\epsilon|x)q_g(n)\alpha_3(x, y)
    \label{eq:keratlas}
\end{align}

where $\mathbbm{1}$ is the indicator variable. The three acceptance probabilities can be estimated by maintaining DB (\cref{eq:db}) for each term (line) in \cref{eq:keratlas} separately, and are derived in \cref{sec:accept_prob}.
\subsection{Warmup}
\label{sec:warmup}
ATLAS needs to tune two parameters in warmup:

\vspace{-8pt}
\paragraph{Baseline step size $\epsilon_0$:} For this, we simulate small HMC trajectories (e.g, 20 leapfrog steps) and use dual averaging adaptation~\citep{hoffman2014no} to target an acceptance rate of $60-65\%$.
\vspace{-8pt}
\paragraph{Distribution of trajectory lengths $q_g(n)$:} For this, we run a short chain with NoUT proposals with step size $\epsilon_0$, and save the U-turn trajectory lengths ($n_\rmut$) for every iteration. 
We use this collection of trajectory lengths to construct $q_g(n)$.
In this work, we construct a uniform distribution between the 10th and 90th percentile of u-turn lengths.
ATLAS is not very sensitive to this tuning as the trajectories from $q_g(n)$ are only used in the DR-upon-failure proposals.

\section{Numerical Experiments}
\label{sec:experiments}

\begin{figure*}[t]
\centering
\begin{subfigure}[b]{\textwidth}
   \includegraphics[width=1\linewidth]{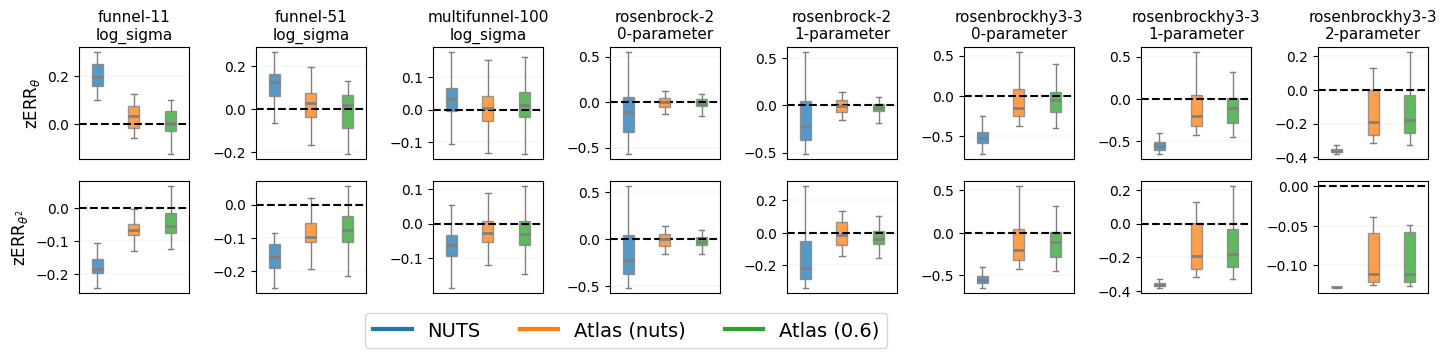}
   \caption{We show the normalized error in parameters and squared parameters for NUTS (blue), ATLAS with the same baseline step size as NUTS (in orange), and with baseline step size tuned to target acceptance rate of 0.6 (green).}
   \label{fig:boxplot-complex}
\end{subfigure}
\vspace{5pt}
\begin{subfigure}[b]{\textwidth}
   \includegraphics[width=1\linewidth]{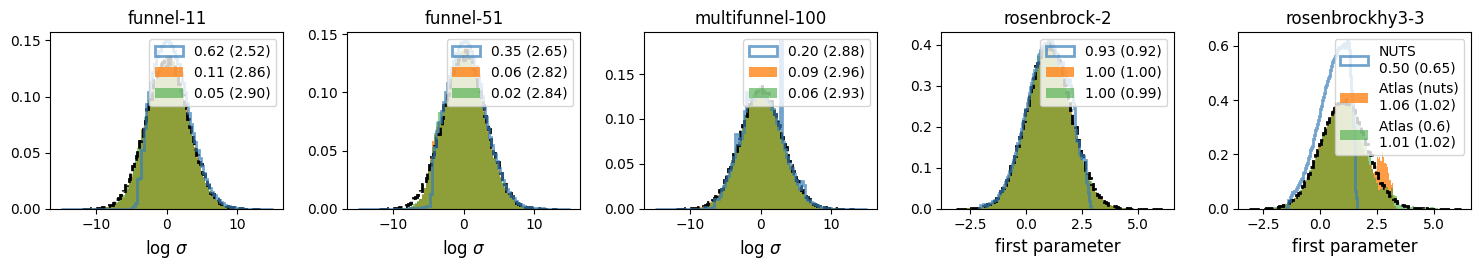}
   \caption{Corresponding histograms of the $\log \sigma$ parameter for funnel and the first parameter in Rosenbrock distribution. Their true distribution is shown in black dashed lines, and is $\mathcal{N}(0, 3)$ and $\mathcal{N}(1,1)$ respectively. NUTS results are shown in red line and ATLAS results are two filled histograms. Legends show the inferred mean (standard deviation).}
   \label{fig:hist-complex}
\end{subfigure}
\vspace{-20pt}
\caption[Atlas]{Models with complex geometry, e.g., variants of Neal's funnel and Rosenbrock distribution. NUTS with fixed step size fails to correctly sample these distributions, but ATLAS with step size adaptation is able~to.
}
\label{fig:atlas}
\end{figure*}
\vspace{-5pt}

\begin{figure*}[ht]
    \centering
    \includegraphics[width=0.99\textwidth]{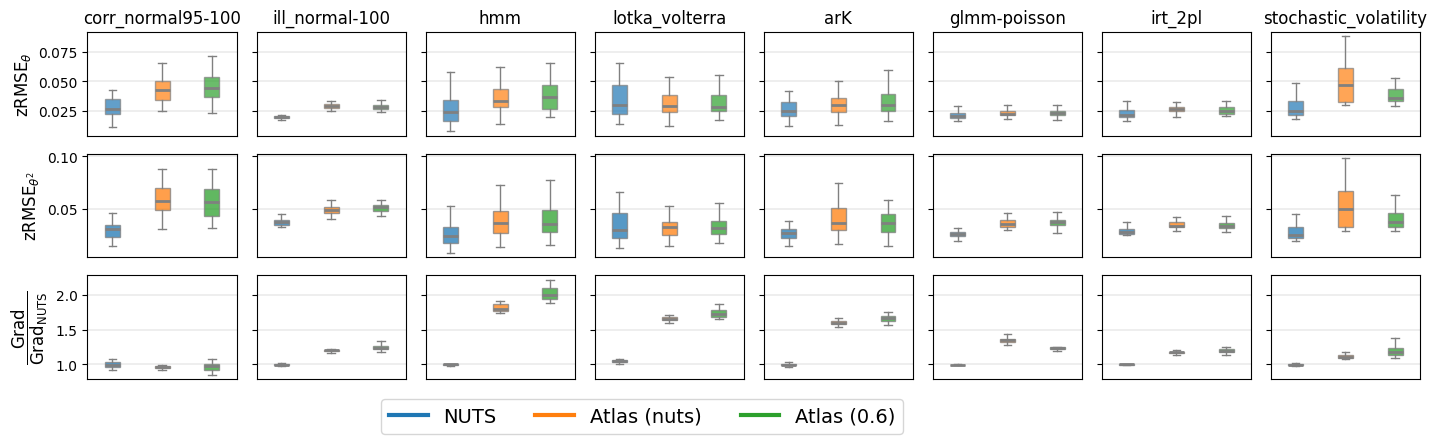}
    \caption{Normalized RMSE for baseline models where NUTS is accurate. Third row shows the cost in terms of number of gradient evaluations normalized against NUTS. ATLAS is computationally competitive to~NUTS.}
\label{fig:boxplot-allexp}
\end{figure*}

\begin{figure*}
    \centering
    \includegraphics[width=0.99\textwidth]{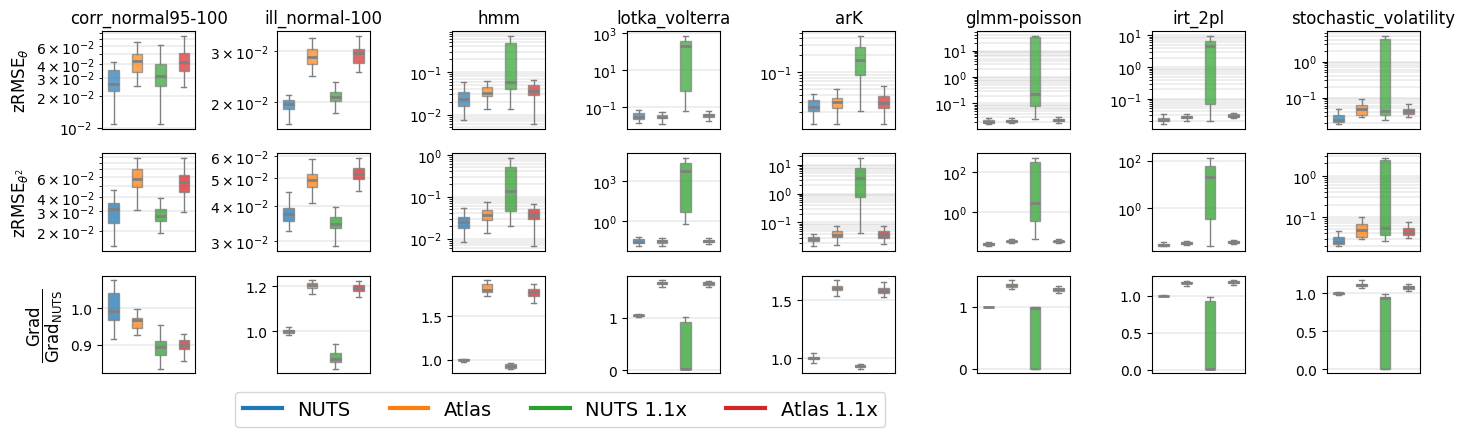}
    \caption{Normalized RMSE for NUTS (blue), and when the step size is increased by 10\% (NUTS 1.1x in green), as well as ATLAS for the same two step sizes. NUTS with larger step sizes fails, but ATLAS does not}
\label{fig:boxplot-stepsize}
\end{figure*}

We now compare the performance of ATLAS against NUTS as implemented in \texttt{cmdstanpy}~\citep{hoffman2014no, carpenter2017stan}.

\vspace{-5pt}
\paragraph{Models Evaluated:}
We will show results for two different suites of models. 
We use `complex' models that have varying curvature to illustrate the necessity of step size adaptation. These consist of- 
a 11 and 51-D (dimensional) Neal's funnel (\texttt{funnel-11, funnel-51}), 
10 independent copies of 10-D Neal's funnel (\texttt{multifunnel-100}),
a 2-D Rosenbrock (\texttt{rosenrbock-2}), and
a more challenging 3-D hybrid Rosenbrock~\citep{Pagani19-rosenbrock} (\texttt{rosenrbockhy3-3}).
The reference (true) samples for these models are generated analytically.

In addition, we also evaluate `baseline' models where NUTS is accurate to show that ATLAS is competitive with NUTS for simpler distributions. These include-
a 100-D correlated Gaussian with correlation coefficient $r=0.95$ 
(\texttt{corr\_normal95-100}),
a 100-D ill-conditioned Gaussian (\texttt{ill\_normal-100} ),
4-D hidden Markov model (\texttt{hmm}),  
8-D Lotka-Volterra population dynamics  (\texttt{lotka\_volterra}),
7-D autoregressive time series (\texttt{arK}),
a 45-D Gaussian linear model (\texttt{glmm-poisson},
a 143-D item-response theory model (\texttt{irt\_2pl}),
and a 503-D stochastic volatility model (\texttt{stochastic\_volatility}).
We generate the reference samples for these models by running 16 chains of NUTS for 100,000 samples each. The step size is tuned to target $95\%$ acceptance, and we adapt a dense mass matrix in the warmup phase for 5000 iterations. 

\vspace{-5pt}
\paragraph{Experimental Setup:} 
We run 32 chains for 2000 samples each for baseline models, and 50000 samples for complex models.
The step size for NUTS is tuned for 1000 iterations to target the $80\%$ acceptance during warmup.
We fix mass matrix to be identity.

For ATLAS, we use default hyperparameters discussed in \cref{sec:stepadapt} and \ref{sec:atlas}.
We show results for two baseline step sizes $\epsilon_0$- one same as NUTS and another fit during 100 warmup iterations to target $60\%$ acceptance.
The warmup phase for constructing the global distribution of trajectory lengths $q_g(n)$ lasts for 100 proposals.

\vspace{-5pt}
\paragraph{Metric for comparison:}

For each chain, we evaluate normalized error for parameters ($\theta$) and squared-parameters ($\theta^2$) along every dimension 
\begin{equation}
 \mathrm{zERR}_{\theta_d} = \cfrac{\big(\frac{1}{N}\sum_i\theta_{d}^{(i)}\big) - \mu(\theta_{\mathrm{ref},d}) }{\sigma(\theta_{\mathrm{ref},d})} 
\end{equation}
\begin{equation}
 \mathrm{zERR}_{\theta^2_d} = \cfrac{\big(\frac{1}{N}\sum_i(\theta_{d}^{(i)})^2\big) - \mu(\theta^2_{\mathrm{ref},d}) }{\sigma(\theta^2_{\mathrm{ref},d})} 
\end{equation}
where $\mu(\theta_{\mathrm{ref},d})$ and $\sigma^2(\theta_{\mathrm{ref},d})$ are the mean and variance evaluated using reference samples (and similarly for  $\theta_d^2$).
For complex models, we show the error for selected dimensions.
For baseline models, we combine these to evaluate normalized root mean square error 
\begin{equation}
 \mathrm{zRMSE}_{\theta} = \Bigg(\frac{1}{D}\sum_{d=1}^D \mathrm{zERR}_{\theta}^2 \Bigg)^{\frac{1}{2}}
\end{equation}
and similarly for parameters squared ($\theta^2$).
All results are shown as a boxplot across all 32 chains.

\subsection{Accuracy for complex models}    

We begin with the results for complex models in \cref{fig:boxplot-complex}. For Neal's funnel and its variants, we only show the error in log-scale parameter as the latent parameters are sampled accurately with both NUTS and ATLAS. For 2-D and 3-D Rosenbrock, we show the error in all parameters. 
\Cref{fig:hist-complex} shows the corresponding marginal histograms for log-scale parameter for funnels, and the first parameter for Rosenbrocks. 
The legends indicate estimated mean and variance. As can be seen, NUTS is not able to sample these distributions correctly, while ATLAS is. It  underestimates the variance for the third parameter of \texttt{rosenbrockhy3-3}, but is still significantly better than NUTS. 

The boxplots do not show outliers to maintain clarity but NUTS generally has more outlier chains than ATLAS. This is visible in the histogram for \texttt{multifunnel-100} in~\cref{fig:hist-complex} where the histogram for NUTS (red line) shows spikes. 
In other histograms, NUTS clearly misses the tails of the distributions. 

The computational cost of ATLAS for these experiments is approximately 5x, 2x, 1.5x, 5x, 10x more than NUTS for \texttt{funnel-11}, \texttt{funnel-51}, \texttt{multifunnel-100}, \texttt{rosenbrock-2}, \texttt{rosenbrockhy3-3} respectively.
For context, this is similar to the cost of NUTS with a smaller step size, tuned to target $0.95\%$ acceptance (instead of 0.8). However NUTS is unable to sample most of these models \textit{even with this smaller step size}.


\vspace{-3pt}
\subsection{Competitive on baseline~models}
\vspace{-3pt}

\Cref{fig:boxplot-allexp} compares the performance of ATLAS and NUTS on baseline models where the latter is able to sample the distributions correctly.
The first two rows show the normalized RMSE, and  the accuracy of ATLAS is broadly similar to that of NUTS.

In the third row, we compare the computational cost in terms of number of gradient evaluations normalized against average cost of a NUTS chain.
For low dimensional models (\texttt{hmm}, \texttt{lotka\_volterra}, and \texttt{arK}), ATLAS is 1.5x-1.8x more expensive than NUTS, though these models are computationally cheap themselves. 
For larger (and computationally intensive) models like high dimensional Gaussians, \texttt{glmm-poisson} (45-D), \texttt{irt\_2pl} (143-D), and \texttt{stochastic\_volatility} (503-D), ATLAS is only 1.1x-1.25x more expensive.
Hence we conclude that ATLAS is computationally competitive to NUTS for simpler distributions.

\subsection{Robustness to hyperparameters}

Lastly, we study the sensitivity of different algorithms to tuning of parameters in the warmup phase. ATLAS tunes the distribution of trajectories $q_g(n)$ and baseline step size for first proposal $\epsilon_0$, while NUTS tunes only the step size\footnote{We fix the mass matrix to be identity and studying the tuning of mass matrix is out of the scope of this work.}. 
We find that ATLAS is broadly insensitive to fitting $q_g(n)$ since it is used only in the `DR-upon-failure' proposal. Furthermore, if necessary, we can also pool the trajectory lengths from all chains to construct a more robust distribution $q_g(n)$. 

In \cref{fig:boxplot-stepsize}, we study the robustness for tuning the baseline step size. We show the results on baseline models for default configuration of NUTS (in blue) when the step size is tuned to target 80\% acceptance rate, and when this step size is increased by 10\% (in green). ATLAS results are shown for the same two as baseline step size $\epsilon_0$. 
NUTS with larger step size diverges and is unable to sample accurately even from the baseline models, highlighting its sensitivity to hyperparameters like step size~\citep{Livingstone19}. Tuning step size in warmup can fail for few reasons, e.g., if the warmup phase is not long enough, initialization is poor or the sampler is unable to explore a region due to complex geometry. 
On the other hand, due to its ability to adapt step size, the accuracy of ATLAS is not very sensitive to this tuning.
The results remain the same for Neal's funnel and Rosenbrock distributions.

\vspace{-7pt}
\section{Discussion and Future Work}
\label{sec:discussion}
\vspace{-7pt}


We have presented a strategy to locally adapt the step size parameter of HMC, and combined it with a simple no U-turn strategy to adapt trajectory lengths to develop an adaptive sampler, ATLAS:~\emph{adapting trajectory length and step-size}.
We have shown that unlike NUTS with fixed step size, ATLAS is able to sample complex distributions with varying curvature like Neal's funnel and Rosenbrock distribution. It is also robust to tuning of hyperparameters like baseline step size during warmup. 
To make ATLAS computationally efficient and competitive with state-of-the-art samplers like NUTS, we implemented it in a delayed-rejection framework. Here special care was required to marginalize over multiple possible proposals of NoUT trajectory as the first proposal map. 
Overall, ATLAS presents a robust alternative to NUTS.

There are many avenues for future work. We have experimented with a few choices for the form of step size distribution (lognormal), strategies to approximate Hessian and maximum eigenvalues, but these can be improved upon with more experiments. For trajectory adaptation, we used No-U-Turn sampler based on \cite{Bou-Rabee24} for its simplicity, but it has inefficiencies like sub-u-turns. Combining step size adaptation with more efficient adaptive samplers like NUTS can improve results. Finally, we have chosen identity mass matrix in this work, but it affects the performance of HMC very strongly. It is worth studying if our approach of approximating the local Hessian can be used to adapt local mass matrix~\citep{girolami2011riemann}.



\bibliography{ref, ref2}
\onecolumn
\appendix
\appendixpage
\numberwithin{equation}{section}
\numberwithin{figure}{section}
\numberwithin{table}{section}
\title{Supplementary Material}
\section{Notation}
\label{sec:notation}

We begin by setting up the notation that will be used in the algorithms throughout this supplementary section. 

We seek to generate samples from a target probability density function $\pi(\theta)$ with parameters $\theta \in \mathbb{R}^D$.
We use $g$ to refer the gradient of the target $g=\nabla_\theta \pi(\theta)$. 

In Hamiltonian Monte Carlo (HMC), $\rho$ represents auxiliary momentum. $x$ denotes the state in the phase space for HMC, i.e.,  $x=(\theta, \rho)$.
We will use $\tilde \pi$ to denote the Gibbs density which is the combined density of $(\theta, \rho)$.
$\mathcal{F}$ is the flip operator that switches the direction of the momentum, so $\mathcal{F}(x)=(\theta, -\rho)$.
$\mathcal{L}^n_\epsilon$ is the operator denoting leapfrog trajectory simulated with step size $\epsilon$ for $n$ steps. So the forward map of HMC can be represented as $y=\mathcal{FL}^n_\epsilon(x)$.

We will use `$\tilde{}$' to denote list of points sampled on a HMC trajectory, i.e., $\tilde{\theta}$ and $\tilde{g}$ is a list of $\theta$ and $g$ sampled in an HMC trajectory. $i^{th}$ element of this list, e.g., $\tilde{x}$, is denoted as $x^{(i)}$.

\begin{algorithm}[H]
    \DontPrintSemicolon
    \SetKwInOut{KwG}{Globals}
    \SetKwInOut{KwIn}{Input}
    \SetKwInOut{KwOut}{Output}

    \KwG{Target distribution $\pi$,
    Mass matrix $M$}
    \KwIn{Initial state $x$,  step size~$\epsilon$, number of steps $n$}
    \KwOut{State after $n$ leapfrog steps}

    \vspace{\baselineskip}
    $\theta^{(0)}, \rho^{(0)} = x$ \;
    \For{$i$ in $1 \ldots n$}
    {
    	$\rho^\prime \leftarrow \rho^{(i)} + \frac{\epsilon}{2} \nabla \log \pi(\theta)|_{\theta^{(i)}}$ \;
    	$\theta^{(i+1)} \leftarrow \theta^{(i)} + \epsilon M^{-1} \rho^\prime$ \;
    	$\rho^{(i+1)} \leftarrow \rho^\prime + \frac{\epsilon}{2} \nabla \log \pi (\theta)|_{\theta^{(i+1)}}$ \;
     }
	\Return{$(\theta^{(n)}, \rho^{(n)})$}
\caption{Leapfrog integration  (\texttt{Leapfrog})}
\label{alg:leapfrog}
\end{algorithm}

\section{NoUT sampler for trajectory adaptation}
\label{sec:nout_details}

We begin by presenting the algorithm for sampling a No U-Turn trajectory described in \cref{alg:nouturn} that was discussed in \cref{sec:nout}.
For a given initial state and step size $\epsilon$, it returns a list of states ($\tilde{x}$) and gradients ($\tilde{g}$) visited along the trajectory, as well as the number of steps taken to make a U-turn ($n_\rmut$). 

In \cref{alg:nouturnsampler}, we use this to build a simple No U-Turn sampler with adaptive trajectory lengths and fixed step size $\epsilon_0$, that was originally presented in \cite{Bou-Rabee24}.
Note that to evaluate acceptance probability, we need to simulate a no u-turn trajectory in the reverse direction from the proposed state. 
We assume that we can store the number the states visited in the forward trajectory, so that the U-turn condition in the reverse trajectory for these points can be evaluated without any extra computation. 

\begin{algorithm}
    \DontPrintSemicolon
    \SetKwInOut{KwG}{Globals}
    \SetKwInOut{KwIn}{Input}
    \SetKwInOut{KwOut}{Output}
    \SetKwComment{Comment}{$\triangleright$\ }{}
    \SetKwFunction{LeapFrog}{LeapFrog}

    \KwG{Target distribution $\pi$,
    Mass matrix $M$
    }
    \KwIn{Initial state $x$,
    step size $\epsilon$, 
    maximum number of steps $N_{\rmmax}=1024$
    }
    \KwOut{Return trajectory up-to U-turn. 
    }

    \vspace{\baselineskip}
    $\theta^{(0)}, \rho^{(0)}= x$; $\quad$   
    $g^{(0)} = \nabla_\theta \pi|_{\theta^{(0)}}$\;
    $\tilde{\theta},\, \tilde{\rho},\, \tilde{g} = [\theta^{(0)}],\, [\rho^{(0)}],\, [g^{(0)}]$\;
    $d = 0$ \hfill \tcp*[f]{Initialize distance to zero}\;
    \For{$j$ in $1 \ldots N_{\rmmax}$}{
        $x^{(j)}$ = \LeapFrog{$x^{(j-1)}, \epsilon, 1$}\;
        $\theta^{(j)}, \rho^{(j)} = x^{(j)}$; 
        $\quad g^{(j)} = \nabla_\theta\pi|_{\theta^{(j)}}$\;
        $\tilde{\theta}.\mathrm{prepend}(\theta^{(j)})$; 
        $\tilde{\rho}.\mathrm{prepend}(\rho^{(j)})$;
        $\tilde{g}.\mathrm{prepend}(g^{(j)})$\;        
        \lIf(\tcp*[h]{Check no U-Turn condition}){ $||\theta^{(j)} - \theta^{(0)}||_2 > d$} 
        { $d = ||\theta^{(j)} - \theta^{(0)}||_2$}
        \lElse{\textbf{\texttt{break}} \hfill \tcp*[h]{U-Turn made. Stop}}
    }
    $\tilde {x} = [\tilde{\theta},\, \tilde{\rho}]$;\, $n_{\rmut} = j$\;
    \Return $\tilde{x},\, \tilde{g},\, n_{\rmut}$\;
\caption{Trajectory upto U-Turn  (\texttt{TrajUptoUTurn})}
\label{alg:nouturn}
\end{algorithm}

\begin{algorithm}
    \DontPrintSemicolon
    \SetKwInOut{KwG}{Globals}
    \SetKwInOut{KwIn}{Input}
    \SetKwInOut{KwOut}{Output}
    \SetKwComment{Comment}{$\triangleright$\ }{}
    \SetKwFunction{LeapFrog}{LeapFrog}
    \SetKwFunction{TrajUptoUTurn}{TrajUptoUTurn}

    \KwG{Target distribution $\pi$,
    Mass matrix $M$    }
    \KwIn{Initial state $x$,
    step size $\epsilon$, 
    offset from initial position $f_\rmoff$,
    maximum number of steps $N_{\rmmax}=1024$
    }
    \KwOut{Return a proposal point before U-turn, associated lists and distribution.
    }
    \vspace{\baselineskip}
    $\tilde{x}, \tilde{g}, n_\rmut = \TrajUptoUTurn(x, \epsilon_0)$\;
    $q(n|x) = \mathcal{U}(\lfloor n_{\rmut}f_\rmoff\rfloor, n_{\rmut})$ \; 
    $n \sim q(n|x)$; \quad $x' = \tilde{x}^{(n)}$ \hfill \tcp*[h]{Proposal point}\;
    \tcc*[h]{Flip proposal and simulate ghost trajectory to construct $q(n|x')$}
    $\tilde{x'}, \tilde{g'}, n'_\rmut = \TrajUptoUTurn(\mathcal{F}(x'), \epsilon_0)$ \;   
    $q(n|x') = \mathcal{U}(\lfloor n^g_{\rmut}f_\rmoff\rfloor, n^g_{\rmut})$ \;
    $\alpha = \frac{\tilde \pi(x')}{\tilde \pi(x)}\frac{q(n|x')}{q(n|x)}$; \, $u \sim \mathcal{U}(0, 1)$ \hfill \tcp*[h]{Acceptance probability}\;
    \lIf{$u < \alpha$}{   \Return$x'$    }
    \lElse{\Return $x$}
\caption{Simple No U-Turn Sampler  (\texttt{NoUTSampler})}
\label{alg:nouturnsampler}
\end{algorithm}

\section{Strategy for adapting step-size}
\label{sec:stepadapt_details}

\Cref{alg:stepadapt} presents our approach for constructing a step size function given a list of positions ($\tilde{x}$) and gradients ($\tilde g$) along a HMC trajectory. 
For robustness, if the length of the input HMC trajectory is less than a minimum number of samples expected for a good Hessian approximation ($N_H$), or if the power iteration fails, the algorithm retries $N_\rmtry$ times by reducing the current stepsize and simulating a new small HMC trajectory to regenerate the requisite list of positions and gradients.
After $N_\rmtry$ attempts, we set the stable step size to a small, pre-chosen step size and generate the step size distribution. 

\begin{algorithm}
    \DontPrintSemicolon
    \SetKwInOut{KwG}{Globals}
    \SetKwInOut{KwIn}{Input}
    \SetKwInOut{KwOut}{Output}
    \SetKwComment{Comment}{$\triangleright$\ }{}
    \SetKwFunction{HessianApprox}{HessianApprox}
    \SetKwFunction{PowerIteration}{PowerIteration}
    \SetKwFunction{StepSizeDistribution}{GetDistribution}
    \SetKwFunction{LeapFrog}{LeapFrog}
    \SetKwFunction{Size}{Size}
    \SetKwFunction{LogNormal}{LogNormal}
    \SetKwProg{Fn}{Function}{:}{}
    
    \KwG{Target distribution $\pi$,
    Mass matrix $M$    }
    \KwIn{
    List of positions states $\tilde{x}$, 
    list of corresponding gradients $\tilde{g}$, 
    initial step size $\epsilon_0$, 
    maximum reduction in step size  $r=1024$, 
    number of samples to estimate Hessian $N_H=10$, 
    number of attempts $N_\rmtry=10$, 
    Width of stepsize distribution $\log\sigma=
    \log{1.2}$}
    \KwOut{Distribution function of step size and a sample from it: $\epsilon, q(\epsilon|x)$
    }
    
    \vspace{\baselineskip}
    $i,\, \epsilon,\, \epsilon_\rmmin=0,\, \epsilon_0,\, \epsilon_0/r$\;    
    $\tilde{\theta}, \tilde{\rho} = \tilde{x}$ \hfill \tcp{Get list of positions}

    \For{$i$ in $1 \ldots N_\rmtry$}
    {
        \If(\hfill \tcp*[h]{Atleast $N_H$ points}){\Size{$\tilde{\theta}$} $\geq N_H$}
        {
            $\hat{H}$ = \HessianApprox{$\tilde{\theta}, \tilde{g}$} \hfill \tcp*[h]{LBFGS}\;
            $\hat{\lambda}_\rmmax$ = \PowerIteration{$\hat{H}$}; 
             $\epsilon_s = \frac{1}{2\sqrt{\hat{\lambda}_\rmmax}}$\;
            \If{$(\hat{\lambda}_\rmmax > 0)$ \& $(\epsilon_s > \epsilon_\rmmin)$}
            {
                $q(\epsilon|x)$ = \StepSizeDistribution{$\epsilon_s, \log\sigma$} \;                
                \Return $q(\epsilon|x)$
             }
        }
        $\epsilon = \epsilon/2$ \hfill \tcp*[h]{Reduce stepsize \& re-try}\;
        $x^{(0)}, \rho^{(0)} = \tilde x^{(0)}$; $g^{(0)} = \nabla_\theta(\pi)|_{\theta^{(0)}}$\;
        \For{$j$ in $1 \ldots N_H$}{
            $x^{(j)}$ = \LeapFrog{$x^{(j-1)}, \epsilon, 1$}\;
            $\theta^{(j)}, \rho^{(j)} = x^{(j)}$; $\quad$
            $g^{(j)} = \nabla_\theta(\pi)|_{\theta^{(j)}}$\;
        }
        $\tilde{\theta} = [\theta^{(N_H)}, \ldots,\theta^{(0)}]$ \hfill \tcp*[h]{Reverse order for better LBFGS approximation}\;
        $\tilde{g} = [g^{(N_H)}, \ldots,g^{(0)}]$ \;
    }
    $\epsilon_s = 2\epsilon_\rmmin $ \hfill\tcp*[h]{Upon failure, use $\epsilon_\rmmin$ to set $\epsilon_s$}\;
    $q(\epsilon|x)$ = \StepSizeDistribution{$\epsilon_s, \log\sigma$} \;
    \Return $q(\epsilon|x)$

    \vspace{\baselineskip}
    \Fn{\StepSizeDistribution{$\epsilon_s, \log\sigma$}}{
        $\mu = \log(\epsilon_s)-\frac{(\log\sigma)^2}{2}$; 
        $\sigma^2 =  (\log\sigma)^2$ \tcp*[h]{Parameters of corresponding Normal distribution} \;
        $q(\epsilon|x) = $\LogNormal{$\mu, \sigma^2$} \tcp*[h]{This is defined as $\log$ of  $\mathcal{N}(\mu, \sigma^2)$} \;
        \textbf{return}  $q(\epsilon|x)$
    }
    \caption{Step Size Distribution (\texttt{StepSizeDist})}
    \label{alg:stepadapt}
\end{algorithm}

\section{Simple adaptive sampler}
\label{sec:atlas_simple_details}

Here we present more details on the simple adaptive sampler briefly presented in \cref{sec:simpleatlas}. 
In this, every iteration consists of two steps:
(1) construct the step-size distribution and propose a step-size $\epsilon$; then
(2) make a NoUT proposal $y=x^{(n)}$ by simulating a trajectory with this step-size until u-turn. 
A reverse trajectory needs to be similarly simulated to evaluate the acceptance probability. 
The full pseudocode is given in \cref{alg:simplesampler} that combines the trajectory adaptation  and step-size adaptation of \cref{alg:nouturnsampler} and \cref{alg:stepadapt} respectively. 
We will refer to this sampler as ATLAS-Simple. 

\begin{algorithm}
    \DontPrintSemicolon
    \SetKwInOut{KwG}{Globals}
    \SetKwInOut{KwIn}{Input}
    \SetKwInOut{KwOut}{Output}
    \SetKwComment{Comment}{$\triangleright$\ }{}
    \SetKwFunction{GISTProposal}{GISTProposal}
    \SetKwFunction{StepSizeDist}{StepSizeDist}
    \SetKwFunction{LeapFrog}{LeapFrog}

    \KwG{Target distribution $\pi$,
    Mass matrix $M$    }
    \KwIn{Initial state $x$, 
    step size $\epsilon_0$, 
    offset from initial position $f_\rmoff$
    }
    \KwOut{Distribution function of step size at the current point $(\theta,p)$: $f_\epsilon$
    }
    
    \vspace{\baselineskip}
    $\theta, \rho = x$; $\quad$   $g = \nabla_\theta \pi |_\theta$\;
    $q(\epsilon|x)$ = \StepSizeDist{$[x], [g], 2\epsilon_0$} \hfill  \tcp*[h]{This runs short HMC to adapt as $Size([x])=1<N_H$} \;
    $\epsilon \sim q(\epsilon|x)$\;    
    $\tilde{x}, \tilde{g}, n_\rmut = \TrajUptoUTurn(x, \epsilon)$\;
    $q(n|x) = \mathcal{U}(\lfloor n_{\rmut}f_\rmoff\rfloor, n_{\rmut})$ \; 
    $n \sim q(n|x)$; \quad $x' = \tilde{x}^{(n)}$ \hfill \tcp*[h]{Proposal point}\;

    \tcc{Flip proposal and run reverse trajectory to construct $q(\epsilon|x'), q(n|x')$'}
    $\theta', \rho' = x'$; $\quad$   $g' = \nabla_\theta \pi |_{\theta'}$\;
    $q(\epsilon|x')$ = \StepSizeDist{$[\mathcal{F}(x')], [g'], 2\epsilon_0$}\;
    $\tilde{x'}, \tilde{g'}, n'_\rmut = \TrajUptoUTurn(\mathcal{F}(x'), \epsilon)$ \hfill \tcp*[h]{Step size is kept same as forward trajectory} \;   
    $q(n|x') = \mathcal{U}(\lfloor n^g_{\rmut}f_\rmoff\rfloor, n^g_{\rmut})$ \;
    
    \tcc{Acceptance probability and MH step}
    $\alpha = \cfrac{\tilde \pi(x')}{\tilde \pi(x)} \times \cfrac{q(\epsilon|x') q(n|x')}{q(\epsilon|x)q(n|x)}$ \;
    \lIf{$u < \alpha$}{   \Return$x'$    }
    \lElse{\Return $x$}

\caption{Simple Adaptive Sampler (ATLAS-Simple)}
\label{alg:simplesampler}
\end{algorithm}

\begin{figure}[h]
    \centering
    \includegraphics[width=0.99\textwidth]{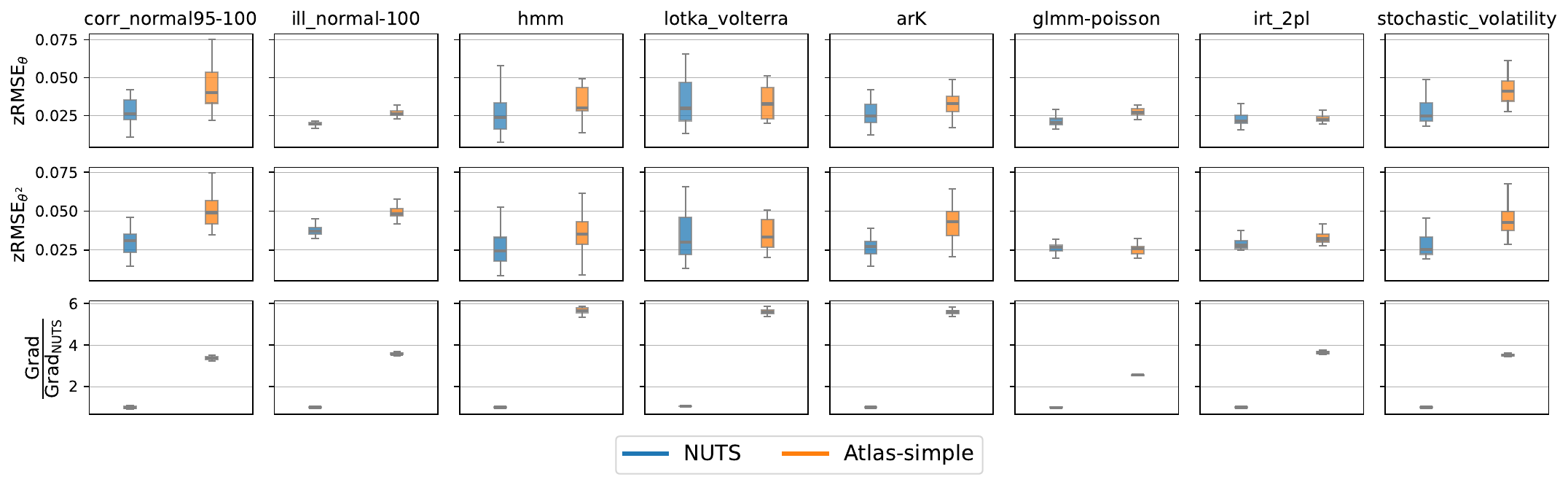}
    \caption{Comparing NUTS (blue) vs simple adaptive sampler of \ref{alg:simplesampler}(orange): First two rows show normalized RMSE for baseline models where NUTS is accurate. Third row shows the cost in terms of number of gradient evaluations normalized against NUTS. ATLAS-Simple is much more expensive that NUTS.}
\label{fig:boxplot-simple}
\end{figure}

As alluded in \cref{sec:simpleatlas}, while this sampler adapts both step-size, and trajectory lengths and is able to robustly sample distributions with complex geometries, it turns out to be very expensive for simple distributions. 
This is shown in \cref{fig:boxplot-simple} where we compare ATLAS-Simple and NUTS on baseline models. 
As in the main text, we ran 32 chains for 2000 samples each for both algorithms. 
The first two rows show the normalized RMSE for parameters ($\theta$) and parameters squared ($\theta^2$), demonstrating that both are able to accurately sample these distributions. 
The last row compares the cost in terms of number of gradient evaluations. ATLAS-Simple is up-to 6x more expensive than NUTS for small models (\texttt{hmm}, \texttt{lotka\_volterra}, and \texttt{arK}), and 2-4x more expensive for larger models. 
On the other hand, due to delayed rejection (DR), ATLAS is only 1.5x-1.8x more expensive than NUTS.
DR makes sure that we adapt the step-size only when necessary. Even then, whenever possible, it re-uses the positions and gradients computed on the rejected trajectory to approximate Hessian and adapt step-size. These make ATLAS in DR framework computationally efficient.

\section{Acceptance Probability for Adaptive samplers for HMC}
\label{sec:accept_prob}

In this section, we derive the acceptance probability for the adaptive samplers based on the forward map of HMC as discussed in this work, and ultimately derive the acceptance probability of various stages of ATLAS sampler. 

We begin by recalling that for a given step-size $\epsilon$ and number of leapfrog steps $n$, 
the HMC forward map $F(x) = \mathcal{FL}_\epsilon^n(x)$ is a deterministic volume-preserving involution, i.e., $F^{-1} = F$ (involution), and if $DF(x)$ is the Jacobian derivative matrix, then its determinant is unity $|\det DF(x)| = 1$ (volume preservation).
We use this to first derive the acceptance probability for a proposal maps of the form
\begin{equation}
    q_F(x, y) = q(n|x) q(\epsilon|x) \delta(y - F(x))
    \label{eq:gen-kernel}
\end{equation}
which encapsulates the proposal for NoUT sampler, sampler with adaptive step-size and ATLAS-simple for different choices of $q(n|x)$ and $ q(\epsilon|x)$.
This result will then help us derive acceptance probabilities for ATLAS. 

\subsection{Metropolis Hastings with a deterministic volume-preserving involution}

The discussion here is adapted broadly from \cite{Modi_2023}, who derive the acceptance probability for HMC where the deterministic kernel is $q_F(x, y) = \delta(y - F(x))$. 
Here we extend that result to the case when we sample $\epsilon$ and $n$ from their own distributions to construct the deterministic mapping $F$ as well.

\begin{lem}
  Let $F$ be a volume-preserving involution with parameters $\epsilon$ and $n$ and  $\pi$ be the target density.
  Then MH with the deterministic proposal kernel $q_F$ given by \eqref{eq:gen-kernel},   with acceptance probability $\alpha$ obeying
  \begin{equation}
  \pi(x)  q(n|x) q(\epsilon|x)  \alpha(x,y) = \pi(y)  q(n|y) q(\epsilon|y)  \alpha(y,x)   \qquad \forall x,y \in S
  \label{eq:alrats}      
  \end{equation}
  has detailed balance with respect to $\pi$.
  \label{lem:map}
\end{lem}

\paragraph{Proof}

Consider Metropolis Hastings transition with transition kernel (up-to a rejection term which is symmetric) using the proposal map  from \eqref{eq:gen-kernel} with $F$ as a volume-preserving involution,
and an acceptance probability obeying \eqref{eq:alrats}.
Then we show that the acceptance probability satisfies the weak form of detailed balance, i.e.,
\begin{equation}    
    \int_A\int_B \pi(x)\alpha(x,y)q_F(x,y)\,\, \textrm{d}xdy \; = \; \int_A\int_B \pi(y)\alpha(y,x)q_F(y,x)\,\, \textrm{d}xdy 
    \label{dbmF}
\end{equation}
for all measurable subsets $A,B \subset S$, where $S$ is the continuous state space, which can be taken as  $\mathbb{R}^{2d}$.
We first substitute \eqref{eq:alrats} into the left-hand side,
then apply the sifting property of the delta function,  which is that $\int_{-\infty}^\infty f(x) \delta(x-x_0) = f(x_0)$, to integrate out $y$.
\begin{align}
    \int_A\int_B \pi(x)\alpha(x,y)q_F(x,y)\,\, \textrm{d}xdy & = 
      \int_A\int_B \pi(x) \alpha(x,y)  q(n|x) q(\epsilon|x)\delta(y-F(x))\, \textrm{d}xdy \nonumber \\
        &= \int_A\int_B \pi(y)\alpha(y,x)  q(n|y) q(\epsilon|y) \delta(y-F(x))\, \textrm{d}xdy
        && \hspace{-60pt} \text{Using \cref{eq:alrats}} \nonumber \\
    &= \int_{B\cap F^{-1}(A)} \pi(F(x))  \alpha(F(x),x) q(n|F(x)) q(\epsilon|F(x)) \, \textrm{d}x
        \nonumber \\
    &=\int_{F(B)\cap A} \pi(y) \alpha(y,F^{-1}(y)) q(n|y) q(\epsilon|y)  \cdot |\det DF(F^{-1}(y))|^{-1} dy \nonumber \\
    && \text{\hspace{-100pt} Using change of variables $y=F(x)$} \nonumber \\
    &=\int_{F^{-1}(B)\cap A} \pi(y)  \alpha(y,F(y)) q(n|y) q(\epsilon|y) \, dy
        && \text{\hspace{-157pt} Using $F^{-1}=F$, and volume preservation} \nonumber \\
    &=\int_A\int_B \pi(y)  \alpha(y,x)  q(n|y) q(\epsilon|y)\delta(x-F(y))\, \textrm{d}xdy \nonumber \\
    &=\int_A\int_B \pi(y)  \alpha(y,x)  q_F(y, x)\, \textrm{d}xdy
\end{align}

Hence \cref{eq:alrats} maintains detailed balance for deterministic kernels of the form \cref{eq:gen-kernel}, where we also sample the parameters of the deterministic map locally. 
Then the most efficient choice of $\alpha$ is
\begin{equation}
    \alpha(x, y) = 1 \wedge \frac{\pi(y) q(n|y) q(\epsilon|y)}{\pi(x)q(n|x) q(\epsilon|x)}
\end{equation}

\subsection{ATLAS}

\begin{figure}
    \centering
    \includegraphics[width=0.9\textwidth]{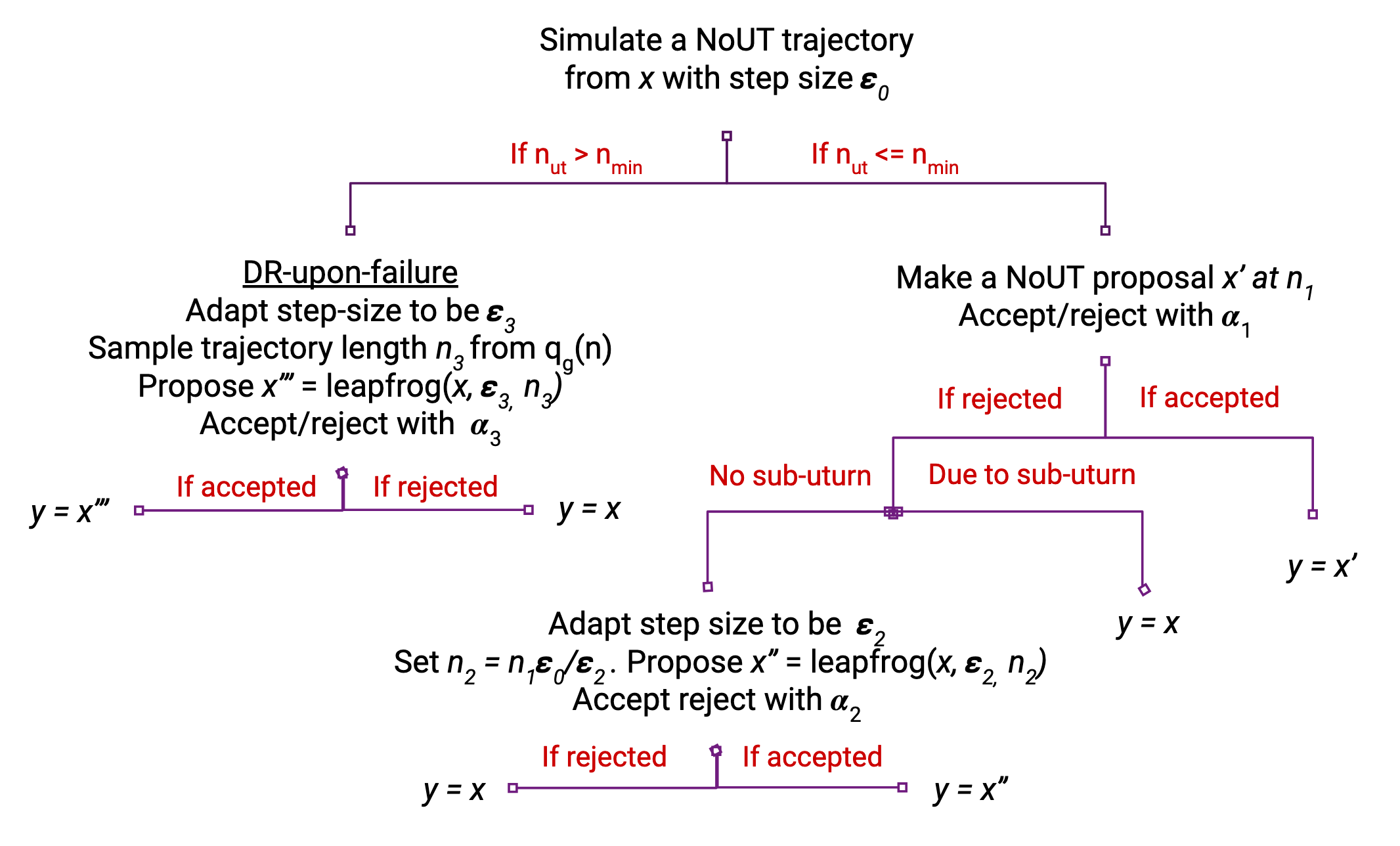}
    \caption{Flowchart of a single iteration of ATLAS represented as a binary tree.}
\label{fig:flowchart}
\end{figure}

The previous result allows us to derive the acceptance probabilities for the various stages of ATLAS. 
The proposal map for ATLAS upto a rejection term is given as
\begin{align}
    q&(x, y)= \mathbbm{1}(n_\rmut > n_\rmmin)\Big[q(n|x) \alpha_1(x, y) \delta(y - F_{1}(x)) +\nonumber  \\
    &\big(1-\mathbbm{1}(subuturn)\big) q(\epsilon|x)q(n|x) [1-\alpha_1(x, F_1(x))] \alpha_2(x, y) \delta(y - F_2(x))\Big] + \nonumber \\
    &\big(1 - \mathbbm{1}(n_\rmut > n_\rmmin)\big)q(\epsilon|x)q_g(n)\alpha_3(x, y)\delta(y - F_3(x))
    \label{eq:keratlas_repeat}
\end{align}
where $F_1 =\mathcal{FL}_{\epsilon_0}^n$, $F_2 =\mathcal{FL}_{\epsilon_2}^{n_2}$ and $F_2 =\mathcal{FL}_{\epsilon_3}^{n_3}$. This is represented as a binary tree flowchart in \cref{fig:flowchart}.
We are interested in deriving the acceptance probabilities $\alpha_1,\, \alpha_2$ and $\alpha_3$. 

We begin by outlining a condition that makes deriving the acceptance probabilities and implementing the ATLAS algorithm quite simple. Note that to construct the distributions $q(n|x')$ and $q(\epsilon|x')$ for evaluating the acceptance probabilities, we need to simulate the reverse trajectories from the proposal point. 
Hence we enforce the following condition--- we evaluate the MH step for any proposal only if we end up in the same branch (of \cref{fig:flowchart}) in the reverse trajectory, as the branch from which the proposal is being made, or else we reject the proposal and fall back to the initial point $x$. 
E.g., if we make a proposal $x''$ in the forward trajectory by rejecting the first proposal without a sub-U-turn, but if the reverse trajectory from $x''$ suffers a DR-upon-failure, we simply reject the proposal $x''$ and do not evaluate the MH criterion. 
This is motivated from the fact that the reversibility of HMC map requires us to sample exactly the same $(\epsilon, n)$ parameters in the reverse trajectory ($x'\rightarrow x$) as were used to make the proposal in the forward trajectory ($x\rightarrow x'$). The probability of being able to sample the same combination of these parameters in different branches, while not zero, is nevertheless exceedingly small as the proposal distributions are constructed differently in every branch.

The assumption of considering the same branch in forward and reverse trajectories makes our criterion more stringent and hence the acceptance probability marginally less efficient than the maximal acceptance probability. 
However it greatly simplifies the implementation of ATLAS algorithm and the form of acceptance probabilities to maintain detailed balance (DB) as we can now treat separate branches individually.
Balancing the same branch with itself also ensures that we correctly account for the indicator functions. 
Hence to evaluate $\alpha_1$, the DB criterion using the result from previous section gives
\begin{align*}
    \pi(x)q(n|x)\alpha_1(x, y) &= \pi(y)q(n|y)\alpha_1(y, x)  \\
    \implies \alpha_1(x, y) &= 1 \wedge \frac{\pi(y)q(n|y)}{\pi(x)q(n|x)}
\end{align*}
If this proposal is rejected, and there is no sub-uturn in the trajectory, we make the delayed proposal with step size adaptation. The DB criterion for it is:
\begin{align*}
    \pi(x)q(\epsilon|x)q(n|x) [1-\alpha_1(x, F_1(x))] \alpha_2(x, y) &= \pi(y)q(\epsilon|y)q(n|y) [1-\alpha_1(y, F_1(y))] \alpha_2(y, x) \\
    \implies \alpha_2(x, y) &= 1 \wedge \frac{\pi(y)q(\epsilon|y)q(n|y) [1-\alpha_1(y, F_1(y))]}{\pi(x)q(\epsilon|x)q(n|x) [1-\alpha_1(x, F_1(x))]}
\end{align*}
Finally, if the first NoUT trajectory already fails, then the DB for the resulting delayed proposal is:
\begin{align*}
    \pi(x)q(\epsilon|x)q_g(n)\alpha_3(x, y) &= \pi(y)q(\epsilon|y)q_g(n)\alpha_3(y, x) \nonumber \\
    \alpha_3(x, y) &= 1 \wedge \frac{\pi(y)q(\epsilon|y)}{\pi(x)q(\epsilon|x)} 
\end{align*}

These are the same acceptance probabilities used in \cref{alg:atlas}, and allow ATLAS to maintain detailed balance.

\end{document}